\documentclass[conf]{new-aiaa}
\usepackage[utf8]{inputenc}
\usepackage{textcomp}
\usepackage{algorithm}
\usepackage{algorithmic}
\usepackage{bm}
\usepackage{array}
\usepackage{graphicx}
\usepackage{amsmath}
\usepackage{float}
\usepackage{amssymb}
\usepackage{subfigure}
\usepackage{xcolor,color,colortbl}
\usepackage[version=4]{mhchem}
\usepackage{siunitx}
\usepackage{longtable,tabularx}
\usepackage{comment}
\title{Adaptive Scale Factor Compensation for Missiles with Strapdown Seekers via Predictive Coding}
\author{Brian Gaudet\footnote{Engineer, E-mail:briangaudet@mac.com}}
\affil{DeepAnalytX, LLC, 1130 Swall Meadows Rd,  Bishop CA 93514}

\begin{document}

\maketitle

\begin{abstract}
In this work we present a method to adaptively compensate for scale factor errors in both rotational velocity and seeker  angle measurements. The adaptation scheme estimates the scale factor errors using a predictive coding model implemented as a deep neural network with recurrent layer, and then uses these estimates to compensate for the error.  During training, the model learns over a wide range of scale factor errors that ideally bound the expected errors that can occur during deployment, allowing the deployed model to quickly adapt in real time to the ground truth error. We demonstrate in a realistic six degrees-of-freedom simulation of an exoatmospheric intercept that our method effectively compensates for concurrent rotational velocity and seeker angle scale factor errors.  The compensation method is general in that it is independent of a given guidance, navigation, and control system implementation.  Although demonstrated using an exoatmospheric missile with strapdown seeker, the method is also applicable to endoatmospheric missiles with both gimbaled and strapdown seekers, as well as  general purpose inertial measurement unit rate gyro compensation.

\end{abstract}

\section{Introduction}
\lettrine{S}{cale} factor measurement errors can have a significant impact on the performance of missiles with strapdown seekers \cite{willman1988effects, shneydor1998missile:1}. These errors take the form $\tilde{x} = (1+\epsilon)x$, where $x$ is the ground truth signal value, $\epsilon$ the scale factor error, and $\tilde{x}$ the measured signal, and in general $\epsilon$ can be a function of some other signal (such as the seeker angles). Whereas a gimbaled seeker can be mechanically stabilized, a strapdown seeker is fixed in the missile body frame. Consequently, if a guidance law requires line of sight measurements in an inertial reference frame (as is the case with proportional navigation \cite{shneydor1998missile:2}), the measurements must be computationally stabilized so that missile body rotations do not result in apparent target acceleration. Specifically, the body frame $B$ seeker azimuth and elevation angles $\theta_u^B$ and $\theta_v^B$ must be rotated back to an inertial reference frame $N$, allowing the transformed seeker angles $\theta_u^N$ and $\theta_v^N$ to be used by the guidance law. This computational stabilization typically requires accurate estimates of the missile's rotational velocity $\boldsymbol\omega$ from rate gyro measurements. Integration of $\boldsymbol\omega$ will then give the change in attitude during the homing phase $\mathbf{dq}$, which can then be used to rotate $\theta_u^B$ and $\theta_v^B$ back to $\theta_u^N$ and $\theta_v^N$ (see Section~\ref{stab_model}). However, due to the scale factor error vector associated with rate gyros used to measure the missile's body rate vector, the rotational velocity observable will actually be $\tilde{\boldsymbol{\omega}} = (1 + \boldsymbol{\epsilon}_{\omega}) \boldsymbol{\omega}$, and the integration of $\tilde{\boldsymbol{\omega}}$ will give a biased estimate of $\mathbf{dq}$, leading to imperfect computational stabilization of the seeker angles $\theta_u^N$ and $\theta_v^N$. The imperfect stabilization implies that  $\theta_u^N$ and $\theta_v^N$  are not actually measured in an inertial reference frame, and the missile body rate will introduce a parasitic component to the measured values of $\theta_u^N$ and $\theta_v^N$ and their time derivatives. 

A second type of scale factor error results from the refraction of incoming electromagnetic waves through a radome or irdome used to protect the seeker in an endoatmospheric application.  Alternately, in an exoatmospheric application, the refraction could be due to lens aberrations in an infrared imaging system. Denoting the refraction angle error as $\theta_r$, we can define the error slopes for the seeker's azimuth and elevation seeker angles as $\epsilon_{\theta_u} = \displaystyle \frac{\partial{\theta_{r}}}{\partial{\theta_u}}$ and $\epsilon_{\theta_v} = \displaystyle \frac{\partial{\theta_{r}}}{\partial{\theta_v}}$ respectively. These error slopes cause the  measured  seeker body frame azimuth and elevation seeker angles to be distorted as $\tilde{\theta}_u^B = (1+\epsilon_{\theta_u})\theta_u^B$ and $\tilde{\theta}_v^B = (1+\epsilon_{\theta_v})\theta_v^B$ \cite{siouris2004missile:3}.  When the refraction is due to the interface between the atmosphere and a radome, $\epsilon_{\theta_u}$ is typically denoted as the radome error slope $R$ in the literature. Now consider a strapdown steerable beam phased array seeker with perfect computational stabilization. As the missile maneuvers according to its guidance law, in general its attitude changes, and the received beam passes through different parts of the radome. Consequently, the measured seeker angles are distorted by the seeker angle scale factor errors $\epsilon_{\theta_u}$ and $\epsilon_{\theta_v}$, introducing a parasitic component into $\tilde{\theta}_u^B$ and $\tilde{\theta}_v^B$. Similar to the case of rotational velocity scale factor errors, this adds a parasitic component to the measured values of $\theta_u^N$ and $\theta_v^N$ and their time derivatives, with the parasitic effect occurring even with a perfectly stabilized seeker.  Although the preceding example dealt with radome refraction, a similar problem could arise  due to lens aberration error. 

The parasitic components of the measured values of $\theta_u^N$ and $\theta_v^N$  due to the scale factor errors creates a parasitic attitude loop \cite{siouris2004missile:1,zarchan2012tactical:5}, which can potentially destabilize the guidance system \cite{nesline1984radome}, leading to large miss distances.  For aerodynamically controlled missiles with radomes, we find that low guidance system time constants, high altitude intercepts, higher scale factor errors, and low missile velocity all increase the impact of the parasitic attitude loop on missile performance \cite{siouris2004missile:1,zarchan2012tactical:5}. For divert thruster controlled exoatmospheric intercepts, the parasitic attitude loop is strengthened by higher scale factor errors and increased center of mass variation during the intercept \cite{gaudet2020_foo}.

The parasitic attitude loop can be attenuated by increasing the guidance system time constant \cite{zarchan2012tactical:5}, but this also reduces the effectiveness of the guidance system. For this reason, there has been considerable interest in developing methods to compensate for scale factor errors, most of which were developed for the application of an endoatmospheric missile with radome. One compensation method applicable to radome scale factor errors is to create a map of radome refraction  error over the entire radome \cite{zarchan1999adaptive}. The map is stored as a table, and depending on where the seeker antennae centerline intersects the radome, the flight computer can compensate for refraction. However, the electrical properties of the radome vary with temperature, leading to significant compensation errors in practice \cite{zarchan1999adaptive}.  It is also possible to reduce radome refraction by intentionally varying the radome thickness during manufacturing \cite{das2005advances}, but again, this may not be completely effective if the radome's electrical properties vary with temperature. 

Active radome refraction compensation approaches include non-destructive dithering of the missile acceleration \cite{zarchan1999adaptive} to allow estimation of the radome slope from bandpass filtered line of sight and body angles. However, the authors acknowledge that the approach would not work for the case where the radome slope varies as a function of seeker angles.  In \cite{yueh1985guidance} the authors use Bayesian inference to estimate a time-varying radome slope using a bank of Kalman filters, each tuned to a one of three ground truth radome slopes, reporting a 20\% to 50\% improvement in miss distance.  The authors of \cite{lin1995radome} use a similar method, but assume the radome slope remains constant during the engagement, reporting small miss distance improvements for compensation within the track loop and more significant improvements for compensation outside the track loop. Finally, in \cite{lin2001stability}, the authors propose using supervised learning to train a neural network to predict seeker angle dependent refraction error, and demonstrate that the compensation is effective in reducing miss distance. 

Most published work on missile scale factor compensation has focused on the simplified planar engagement case and are applicable to missiles with a gimbaled seeker and radome. In contrast, here we present a method to compensate for both seeker angle and rotational velocity scale factor errors in missiles with strapdown seekers.  To our knowledge, this is the first published work describing a method to compensate for both rotational velocity and seeker angle scale factor errors using a realistic strapdown seeker model, with  performance demonstrated in a high fidelity simulator. Our method is completely general in that it does not assume a particular guidance, navigation, and control (GN\&C) system architecture. The method uses an action conditional predictive coding model (PCM). Our PCM is implemented as a deep neural network with a recurrent layer and two linear output heads, the first head predicting the next observation $\mathbf{o}$ and the second head predicting the next scale factor error vector $\boldsymbol{\epsilon} = [\epsilon_{\theta_u}\hspace{5pt}\epsilon_{\theta_v}\hspace{5pt}\boldsymbol{\epsilon_{\omega}}]$. As the model learns to predict future observations, it learns an internal representation that is useful for inferring $\boldsymbol{\epsilon}$ through the error vector's influence on a sequence of observations and actions (see Section~\ref{discussion}).  

\begin{figure}[h!]
\begin{center}
\includegraphics[width=0.8\linewidth]{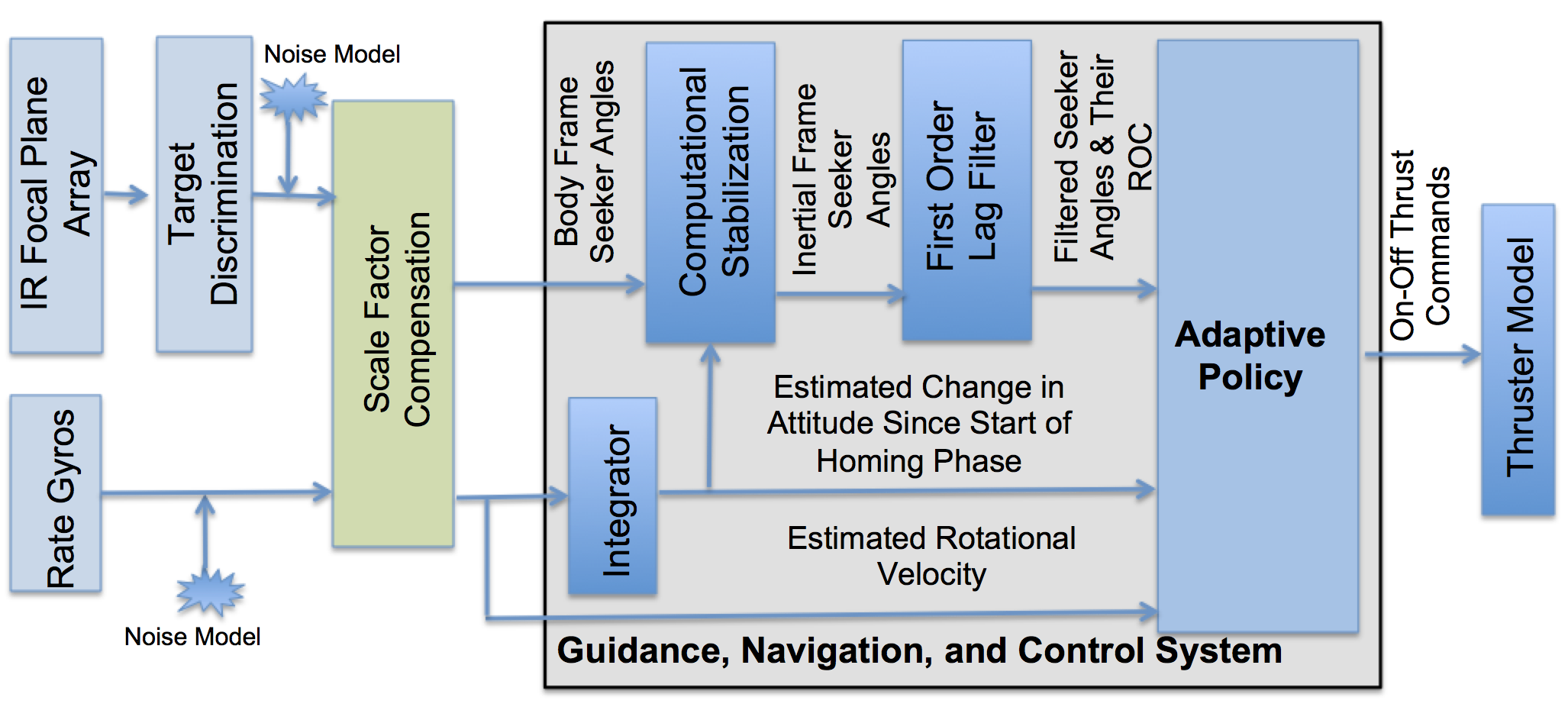}
\caption{System}
\label{fig:System}
\end{center}
\end{figure}

The scale factor compensation  method developed in this work is optimized and tested using the simulator and engagement scenarios described in \cite{gaudet2020_foo}, where an integrated GN\&C system suitable for terminal phase exoatmospheric intercepts against maneuvering targets was optimized using meta reinforcement learning (meta-RL). This high fidelity six degrees-of-freedom (6-DOF) simulator models  parasitic effects including thruster control lag, the parasitic attitude loop resulting from scale factor errors and Gaussian noise on angle and rotational velocity measurements, and a time varying center of mass and inertia tensor caused by fuel consumption and slosh.  The meta-RL optimized GN\&C system was able to adapt in real time to variable environmental and internal dynamics, giving performance close to that of an ideal (no parasitic effects) proportional navigation system with perfect knowledge of the ground truth engagement state. However, the GN\&C system's robustness to constant rotational velocity and angle scale factor errors was limited to the range $-1\times10^{-3} < \epsilon_\theta < 1\times10^{-3}$ and $-1\times10^{-3} < \boldsymbol{\epsilon}_{\omega} < 1\times10^{-3}$.  This was likely due to the inability of the adaptive policy to discriminate between actual target maneuvers and the apparent target maneuvers resulting from the parasitic attitude loop.  Consequently, although the system was robust to small scale factor errors, it could not adapt to larger errors. Our goal in this work is to develop a scale factor compensation method that allows a high probability of successful intercept over a wider range of scale factor errors. A system diagram illustrating the interface between the scale factor compensation block and peripheral system components is given in Fig.~\ref{fig:System}.

The remainder of the paper is organized as follows. Section~\ref{methods} gives background on predictive coding, develops the scale factor compensation method, and describes the PCM network architecture and training approach. Section~\ref{setup} describes the missile configuration, seeker model, equations of motion, engagement scenario, and scale factor error model. Section~\ref{Results} describes the optimization and testing of the scale factor compensation networks, with a discussion of results. 

\section{Methods}\label{methods}

In this section, for some variable with ground truth value $y$, the tilde accent ($\tilde{y})$ denotes the measured value of $y$, the breve accent $\breve{y}$ denotes the estimated value of $y$, and $\bar{y}$ denotes the scale factor error compensated value of $y$. Moreover, $x_{[0:t]}$ denotes values of the variable $x$ ranging from $t=0$ to the current time $t$.

\subsection{Background: Predictive Coding}\label{PCM}

Let $\bf x$ be the ground truth state of an agent interacting with an environment. The environment uses a dynamics model $\mathcal{F}:  \mathbf{x}_{t-1},\mathbf{u}_{t-1} \mapsto \mathbf{x}_{t}$ to update the state of the agent, and the agent  has access to an observation that is a function of the agent's ground truth state $\mathcal{O}: \mathbf{x}_{t} \mapsto \mathbf{o}_{t}$.  Using observation $\bf o$, the agent generates an action $\mathbf{u}$ according to its policy: $\pi: \mathbf{o}_{[0:t]} \mapsto \mathbf{u}_t$. As the agent interacts with the environment, it is possible to learn a predictive model $\mathcal{M}: \mathbf{e}_{0:t}, \mathbf{u}_{0:t} \mapsto \breve{\mathbf{o}}_{t+1}$, where $\mathbf{e}_{t+1}=\breve{\mathbf{o}}_{t+1} - \mathbf{o}_{t+1}$, $\breve{\mathbf{o}}$ is the model's prediction of the next observation, and $\mathbf{e}_0=\mathbf{0}$. This type of predictive model, where the model uses an error signal input to make a prediction, is an example of predictive coding. This is in contrast to models of the type $\mathcal{M}: \mathbf{o}_{t}, \mathbf{u}_{t} \mapsto \breve{\mathbf{o}}_{t+1}$. Note that a predictive coding model's network requires at least one recurrent network layer, as the error $\mathbf{e}_{0:t}$ must be processed over time to estimate $\mathbf{o}_{t+1}$. Predictive coding \cite{rao1999predictive}, was originally developed to explain endstopping in receptive fields of the visual cortex. More recently, in \cite{lotter2016deep}, modern deep learning techniques were applied to predictive coding, demonstrating state of the art results in predicting steering angles from sequential dashboard camera images. Neither of these works used the agent's action as an input to the model, although later in \cite{gaudet2019learning} an action conditional version of predictive coding was used to make  accurate extended predictions of high dimensional trajectories. 

Typically, the PCM networks recurrent layer's hidden state vector is set to zero at the start of an episode $\mathbf{h}_0=\mathbf{0}$. Another approach, first suggested in \cite{gaudet2019learning}, is to learn a mapping from the initial observation in an episode to an initial hidden state vector $\mathcal{H}:\mathbf{o}_0 \mapsto \mathbf{h}_0$.  In certain applications, this can improve the predictive coding model's performance, and we use this technique in this work.

\subsection{Observation Model}\label{obs_model}

Given ground truth missile and target positions $\mathbf{r}_{\mathrm{M}}$ and $\mathbf{r}_{\mathrm{T}}$ in some inertial reference frame $N$, we can define the relative position $\mathbf{r}_{\mathrm{TM}} = \mathbf{r}_{\mathrm{T}} - \mathbf{r}_{\mathrm{M}}$, and we denote the relative inertial frame line of sight unit vector as $\hat{\mathbf{r}}_{\mathrm{TM}}^N = \mathbf{r}_{\mathrm{TM}} / \|\mathbf{r}_{\mathrm{TM}}\|$. Defining $\mathbf{C}_\mathrm{BN}(\mathbf{q})$ as the direction cosine matrix (DCM) mapping from the inertial frame to the body frame given the missile's current attitude $\bf q$, the body frame line of sight unit vector is then computed as $\hat{\mathbf{r}}_{\mathrm{TM}}^B = \mathbf{C}_\mathrm{BN}(\mathbf{q})\hat{\mathbf{r}}_{\mathrm{TM}}^N$.  We can then compute the ground truth body frame target elevation and azimuth angles $\theta_{u}^B$ and $\theta_{v}^B$ as the orthogonal projection of $\hat{\mathbf{r}}_{\mathrm{TM}}^B$ onto the body frame unit vectors  $\hat{\mathbf{u}}=\begin{bmatrix} 0 & 1 & 0 \end{bmatrix}$, $\hat{\mathbf{v}}=\begin{bmatrix} 0 & 0 & 1\end{bmatrix}$, as shown in Equations~\ref{eq:los1} and \ref{eq:los2}. Note that the unit vector corresponding to the unit centerline is $\begin{bmatrix} 1 & 0 & 0\end{bmatrix}$. 

\begin{subequations}
\begin{align}
\theta_{u}^B &= \mathrm{arcsin}(\hat{\mathbf{r}}_{\mathrm{TM}}^B \cdot \hat{\mathbf{u}})\label{eq:los1}\\
\theta_{v}^B &= \mathrm{arcsin}(\hat{\mathbf{r}}_{\mathrm{TM}}^B \cdot \hat{\mathbf{v}})\label{eq:los2}
\end{align}
\end{subequations}

The output of the observation model $\mathcal{O}: \mathbf{x} \mapsto \mathbf{o}$ is then as shown in Equations~\eqref{sf1} through \eqref{sf4}. Here $\boldsymbol{\omega}$ is the ground truth missile rotational velocity vector, $\mathcal{N}(\mu,\sigma,d)$ denotes a $d$ dimensional Gaussian random variable with  mean $\mu$ and standard deviation $\sigma$, keeping in mind that $\epsilon_{\theta_u}$ and $\epsilon_{\theta_u}$ can potentially be functions of the seeker angles $\theta_u$ and $\theta_v$. 

\begin{subequations}
\begin{align}
\tilde{\theta}_u &= (1+\epsilon_{\theta_u}) \theta_u^B + \mathcal{N}(0,\sigma_\theta,1)\label{sf1}\\
\tilde{\theta}_v &= (1+\epsilon_{\theta_v}) \theta_v^B + \mathcal{N}(0,\sigma_\theta,1)\label{sf2}\\
\tilde{\boldsymbol{\omega}} &= (1 + \boldsymbol{\epsilon}_{\omega}) \boldsymbol{\omega} + \mathcal{N}(0,\sigma_\omega,3)\label{sf3}\\
\mathbf{o} &= \begin{bmatrix} \tilde{\theta}_u & \tilde{\theta}_v & \tilde{\boldsymbol{\omega}} \end{bmatrix}\label{sf4}
\end{align}
\end{subequations}

\subsection{Scale Factor Error Model}\label{LADSE}

\begin{figure}[h!]
\begin{center}
\includegraphics[width=0.6\linewidth]{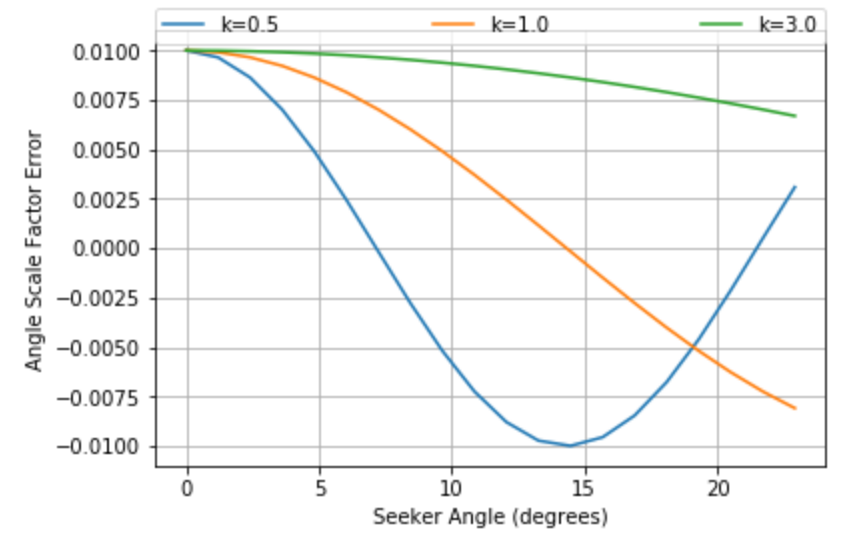}
\caption{Angle Scale Factor as Function of Seeker Angle}
\label{fig:LA_SF}
\end{center}
\end{figure}

During model training, we model the angle scale factor errors $\epsilon_{\theta_u}$ and $\epsilon_{\theta_v}$ as being seeker angle dependent. Although the mechanism of seeker angle dependence differs between exoatmospheric missiles and endoatmospheric missiles with a radome, we use a sinusoidal radome model as suggested in \cite{murray1984correlation}. Although not realistic, this suffices to determine performance with seeker angle dependence.  Specifically, $\epsilon_{\theta_u}$ and $\epsilon_{\theta_u}$ are modeled as shown in Equations~\eqref{la1} and \eqref{la4}, where $k_u$, $k_v$, $\phi_u$, and $\phi_v$ are uniformly distributed random variables sampled at the start of each episode as shown in Table~\ref{tab:asfem}, $A_{\theta_u}$ and $A_{\theta_v}$ determine the maximum magnitude of $\epsilon_{\theta_u}$ and $\epsilon_{\theta_v}$, and $\mathcal{U}(a,b,n)$ denotes an $n$ dimensional uniformly distributed random variable bounded by $(a,b)$, where each dimension of the random variable is independent. Figure~\ref{fig:LA_SF} illustrates the variation of $\epsilon_{\theta_u}$ with $\theta_u$ as a function of $k_u$ with $A_{\theta_u}=1\times10^{-2}$ and $\phi_u=0$. 

\begin{subequations}
\begin{align}
\epsilon_{\theta_u} &= A_{\theta_u} \cos{\left(\frac{2\pi}{k_u}  \theta_u + \phi_u\right)}\label{la1}\\
\epsilon_{\theta_v} &= A_{\theta_v} \cos{\left(\frac{2\pi}{k_v}  \theta_v + \phi_v\right)}\label{la2}\\
A_{\theta_u} &= \mathcal{U}(A_{\theta_u}^\text{MIN},A_{\theta_u}^\text{MAX},1)\label{la3}\\
A_{\theta_v} &= \mathcal{U}(A_{\theta_v}^\text{MIN},A_{\theta_v}^\text{MAX},1)\label{la4}
\end{align}
\end{subequations}

\begin{table}[h!]
	\fontsize{10}{10}\selectfont
    \caption{Angle Scale Factor Error Model Parameter Bounds}
   \label{tab:asfem}
        \centering 
   \begin{tabular}{c  r  r } 
      \hline
      Variable & Lower limit & Upper Limit\\
      $k_u$ & 0.50 & 3.00\\
      $k_v$ & 0.50 & 3.00\\
      $\phi_u$ & $-\pi$ & $\pi$\\
      $\phi_v$ & $-\pi$ & $\pi$\\
      \hline
   \end{tabular}
\end{table}

We also test the scale factor compensation method for the case where $\epsilon_{\theta_u}$ and $\epsilon_{\theta_v}$ are not seeker angle dependent, as shown in Equations~\eqref{a1} and \eqref{a2}. 

\begin{subequations}
\begin{align}
\epsilon_{\theta_u} &= \mathcal{U}\left(A_{\theta_u}^\text{MIN},A_{\theta_u}^\text{MAX},1\right)\label{a1}\\
\epsilon_{\theta_v} &= \mathcal{U}\left(A_{\theta_v}^\text{MIN},A_{\theta_v}^\text{MAX},1\right)\label{a2}
\end{align}
\end{subequations}

The rotational velocity scale factor error vector $\boldsymbol\epsilon_{\omega}$ is sampled from a uniform distribution at the start of each episode and is held  constant throughout the episode, as shown in Equation~\eqref{w1}.

\begin{subequations}
\begin{align}
\boldsymbol\epsilon_{\omega} &=\mathcal{U}\left(-A_{\omega}^{\text{MAX}},A_{\omega}^{\text{MAX}},3\right)\label{w1}
\end{align}
\end{subequations}

\subsection{Scale Factor Compensation Method}\label{sfcm}

Our goal is to optimize a function $\mathcal{G}: \mathbf{o} \mapsto \bar{\mathbf{o}}$ such that the  GN\&C system, using $\bar{\mathbf{o}}$ in place of $\mathbf{o}$, has improved tolerance to scale factor errors $\epsilon_{\theta_u}$, $\epsilon_{\theta_v}$, and $\epsilon_{\boldsymbol{\omega}}$. In this work we take an indirect approach to meeting this goal, and implement $\mathcal{G}$  as shown in Equations~\eqref{eq:comp1} through \eqref{eq:comp4}, where $\breve{\epsilon}_{\theta_u}$, $\breve{\epsilon}_{\theta_v}$, and $\breve{\epsilon}_{\omega}$ are estimated using a predictive coding model (PCM). The observation vector $\bar{\mathbf{o}}$ is then used by the missile GN\&C system (see Fig.~\ref{fig:System}).

\begin{subequations}
\begin{align}
\bar{\theta}_u &= \frac{\tilde{\theta}_u}{1+\breve{\epsilon}_{\theta_u}}\label{eq:comp1}\\
\bar{\theta}_v &= \frac{\tilde{\theta}_v}{1+\breve{\epsilon}_{\theta_v}}\label{eq:comp2}\\
\bar{\boldsymbol{\omega}} &= \frac{\tilde{\boldsymbol{\omega}}}{1+\breve{\epsilon}_{\omega}}\label{eq:comp3}\\
\bar{\mathbf{o}} &= \begin{bmatrix} \bar{\theta}_u & \bar{\theta}_v & \bar{\boldsymbol{\omega}} \end{bmatrix}\label{eq:comp4}
\end{align}
\end{subequations}

\begin{figure}[h!]
\begin{center}
\includegraphics[width=0.5\linewidth]{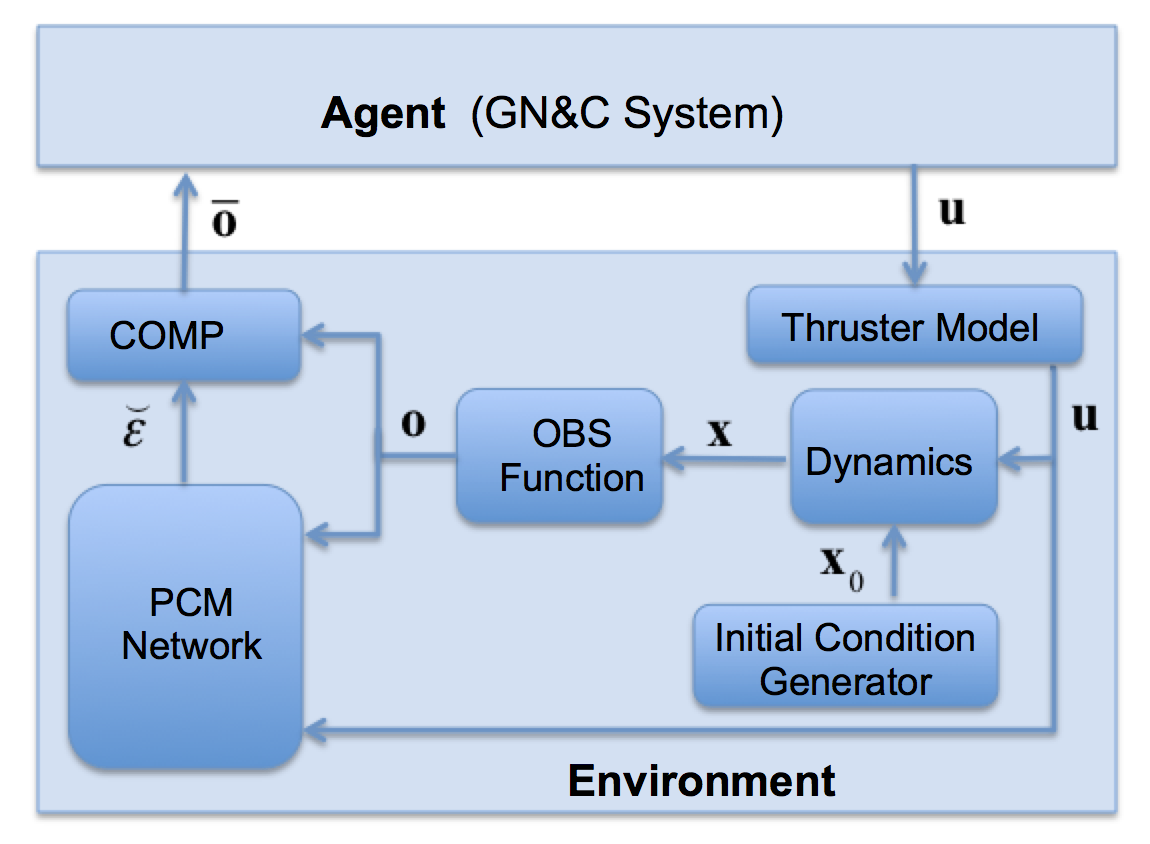}
\caption{Agent - Environment Interface}
\label{fig:AEI}
\end{center}
\end{figure}

The PCM learns as the agent episodically interacts with its environment. The agent-environment interface is shown in Fig.~\ref{fig:AEI}.  Note that the agent is the GN\&C system from Fig.~\ref{fig:System}, with the PCM and scale factor compensation included in the environment. The observation function $\mathcal{O}$ maps the ground truth missile state $\mathbf{x}$ to the observation $\mathbf{o}$ as described in Section~\ref{obs_model}. The episodic interaction between the agent and environment can be modeled as shown in Algorithm~\ref{alg:aei}, where in our application the episode termination condition (done = True) occurs when the target falls outside of the seeker's field of view, the missile exceeds the rotational velocity constraint of 12 rad/s, or the missile runs out of fuel.  At each step of the episode, prior to the dynamics model  $\mathcal{F}$ generating  $\mathbf{x}_{t+1}$, the prediction error  $\mathbf{e}_t$ is added to a buffer $\mathcal{B}_\text{E}$, the recurrent network layer's hidden state $\mathbf{h}$ is added to buffer $\mathcal{B}_\text{S}$, and action $\mathbf{u}=\pi({\mathbf{o}}_{[0:t]})$ is added to buffer $\mathcal{B}_\text{U}$, with the time index $[0:t]$ indicating that the GN\&C system uses a recurrent policy that generates actions based off of the history of observations.  Then, after $\mathcal{M}$ generates  $\mathbf{x}_{t+1}$, the next observation $\mathbf{o}_{t+1} = \mathcal{O}(\mathbf{x}_{t+1})$ is added to a buffer $\mathcal{B}_\text{NEXT\_OBS}$ and the ground truth scale factor error vector $\boldsymbol{\epsilon}=[\epsilon_{\theta_u} \hspace{5pt} \epsilon_{\theta_v} \hspace{5pt} \boldsymbol{\epsilon}_{\omega}]$ is added to the buffer $\mathcal{B}_{\epsilon}$.  These five buffers contain data from the most recent 360 episodes; in the following we will refer to these buffers as the \textit{rollouts}. Note that the buffers are only used for training, not for the deployed scale factor error compensation system.

\begin{algorithm}[h!]
   \caption{Episodic Interaction between Environment and Agent}
   \label{alg:aei}
\begin{algorithmic}
    \STATE Environment: Set $t=0$, $\breve{\boldsymbol{\epsilon}}_t=\mathbf{0}$, $\mathbf{e}_t=\mathbf{0}$, done = False, initialize $\mathbf{x}_t$, generate $\mathbf{o}_t=\mathcal{O}(\mathbf{x}_t)$
    \WHILE{not done}
    \STATE {Agent: generate $\mathbf{u}_t=\pi(\mathbf{o}_{[0:t]})$}
    \STATE {Environment: $\breve{\mathbf{o}}_{t+1}$, $\breve{\boldsymbol{\epsilon}}_{t+1} = \mathcal{M}(\mathbf{e}_t$, $\mathbf{u}_t$)}
    \STATE {Environment: Add $\mathbf{u}_t$ to $\mathcal{B}_\text{U}$ (\textit{Training only})}
    \STATE {Environment: Add $\mathbf{e}_t$ to $\mathcal{B}_\text{E}$ (\textit{Training only})}
    \STATE {Environment: Add $\mathbf{s}_t$ to $\mathcal{B}_\text{S}$ (\textit{Training only})}
    \STATE {Environment: generate $\mathbf{x}_{t+1} = \mathcal{F}(\mathbf{x}_{t},\mathbf{u}_{t})$ and $\mathbf{o}_{t+1}=\mathcal{O}(\mathbf{x}_{t+1})$  }
    \STATE {Environment: Add $\mathbf{o}_{t+1}$ to $\mathcal{B}_\text{NEXT\_OBS}$ (\textit{Training only})}
    \STATE {Environment: Add ${\boldsymbol{\epsilon}}_{t+1}$ to $\mathcal{B}_{\epsilon}$ (\textit{Training only})}
    \STATE {Environment: $\mathbf{e}_{t+1}$ = $\breve{\mathbf{o}}_{t+1}-\mathbf{o}_{t+1}$}
    \IF{$\mathbf{x}_{t+1}$ is terminal}
    \STATE{done = True}
    \ENDIF
    \STATE {Environment: $t=t+1$}
    \ENDWHILE
\end{algorithmic}
\end{algorithm}

\begin{figure}[h!]
\begin{center}
\includegraphics[width=0.6\linewidth]{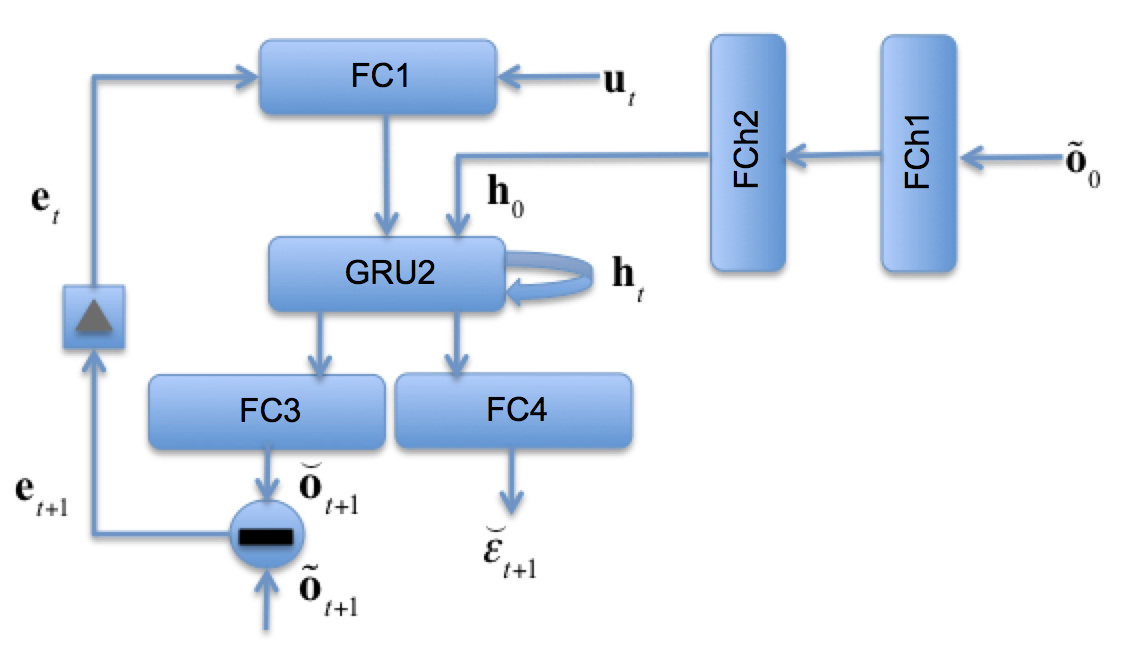}
\caption{PCM Network}
\label{fig:PCM}
\end{center}
\end{figure}

The PCM is implemented as a multi layer neural network with two output heads as shown in Fig.~\ref{fig:PCM}.  where $\text{FC}$ denotes a fully connected layer and $\text{GRU2}$ is a gated recurrent layer \cite{chung2015gated}. $\text{FC1}$, $\text{FCh1}$, $\text{FCh2}$, $\text{FC3}$, and $\text{FC4}$ are implemented as fully connected layers as shown in Equation~\eqref{fc1}, where $\mathbf{W}$ and $\mathbf{b}$ are the weight matrix and bias vector, respectively.  The outputs of $\text{FCh1}$, $\text{FCh2}$, and $\text{FC1}$ are passed through tanh activation functions.

\begin{equation}
\label{fc1}
\mathbf{y} = \mathbf{W}_{xy}\mathbf{x} + \mathbf{b}_{xy}
\end{equation}

The $\text{GRU2}$ layer is implemented as shown in Equations~\eqref{gru1} through \eqref{gru4}, where $\mathbf{W}$ and $\mathbf{b}$ are parameter matrices and vectors, $\mathbf{x}_t$ is the layer input at time $t$, $h$ is the hidden state vector (and also the layer output), $\circ$ denotes a Hadamard product, and $\sigma$ the sigmoid function. The hidden state $\mathbf{h}$ allows the $\text{GRU2}$ layer to learn temporal dependencies in an input data sequence.  Layers $\text{FCh1}$ and $\text{FCh2}$ are used to learn a mapping from the first observation in an episode to an initial value for the recurrent layer's hidden state $\mathcal{H}:\mathbf{o}_0 \mapsto \mathbf{h}_0$. We found this improved performance (increased kill probability by a few percent) as compared to initializing the recurrent layer's hidden state to zeros $\mathbf{h}_0 = \mathbf{0}$.

\begin{subequations}
\begin{align}
\mathbf{r}_t &= \sigma(\mathbf{W}_{xr}\mathbf{x}_t + \mathbf{b}_{xr} + \mathbf{W}_{hh}\mathbf{h}_{t-1} + \mathbf{b}_{hh})\label{gru1}\\
\mathbf{z}_t &= \sigma(\mathbf{W}_{xz}\mathbf{x}_t + \mathbf{b}_{xz} + \mathbf{W}_{hz}\mathbf{h}_{t-1} + \mathbf{b}_{hz})\label{gru2}\\
\mathbf{n}_t &= \tanh{(\mathbf{W}_{xn}\mathbf{x}_t + \mathbf{b}_{xn} + \mathbf{r}_t \circ (\mathbf{W}_{hn}\mathbf{h}_{t-1}+\mathbf{b}_{hn})}\label{gru3}\\
\mathbf{h}_t &= (1-\mathbf{z}_t) \circ \mathbf{n}_t + \mathbf{z}_t \circ \mathbf{h}_{t-1}\label{gru4}
\end{align}
\end{subequations}

\subsection{Training}

The PCM output head $\text{FC3}$ attempts to predict the next measured observation $\mathbf{o}_{t+1}$ by learning a linear mapping from the hidden state $\mathbf{h}_{t}$ to $\mathbf{o}_{t+1}$, whereas $\text{FC3}$ attempts to predict the next ground truth scale factor error vector $\boldsymbol{\epsilon}_{t+1}$ by learning a linear mapping from the hidden state $\mathbf{h}_{t}$ to $\boldsymbol{\epsilon}_{t+1}$.  During training, every 120 episodes $\mathcal{B}_\text{E}$, $\mathcal{B}_\text{S}$, and $\mathcal{B}_\text{U}$ are run forward through the network, generating predictions $\breve{\mathcal{B}}_\text{NEXT\_OBS}$ at the output of $\text{FC3}$ and $\breve{\mathcal{B}}_\epsilon$ at the output of $\text{FC4}$.  The loss $\mathcal{L}$ is then calculated using the cost functions given in Equations~\eqref{cost1} through \eqref{cost3}.

\begin{subequations}
\begin{align}
\mathcal{L}_o &= \sum(\breve{\mathcal{B}}_\text{NEXT\_OBS} - \mathcal{B}_\text{NEXT\_OBS})^2\label{cost1}\\
\mathcal{L}_\epsilon &= \sum(\breve{\mathcal{B}}_\epsilon - \mathcal{B}_\epsilon)^2\label{cost2}\\
\mathcal{L} &= \mathcal{L}_o + \mathcal{L}_\epsilon\label{cost3}
\end{align}
\end{subequations}

During the backward pass, the loss $\mathcal{L}$ propagates backward through the network's computational graph, allowing local computation of the gradients $\nabla_\mathcal{L}\mathbf{W}$ and $\nabla_\mathcal{L}\mathbf{b}$ at each layer.  For the $\text{FC1}$ and $\text{GRU2}$ layers, the input data in the backward pass is unrolled for 60 steps to allow calculating the gradients using the backpropagation through time algorithm \cite{williams1995gradient}, whereas the other layers use the standard backpropagation algorithm \cite{rumelhart1986learning}. The weights are then adaptively updated using the ADAM algorithm \cite{kingma2014adam}. 

Although not shown in Fig.~\ref{fig:PCM}, the correct processing of the recurrent layer's hidden state $\mathcal{B}_\text{S}$ when unrolling the network for the forward pass is critical to learning temporal dependencies between the inputs $\mathbf{e}$, $\mathbf{u}$, and the outputs $\breve{\mathbf{o}}$ and $\breve{\boldsymbol{\epsilon}}$. The reader is referred to \cite{gaudet2019adaptive} for the implementation details in the context of learning a recurrent policy using reinforcement meta-learning.

\section{Experimental Setup}\label{setup}

The scale factor compensation system is optimized and tested using the missile configuration and engagement scenario described in \cite{gaudet2020_foo}. To better understand how the scale factor compensation system interacts with the missile's GN\&C system, we provide  descriptions of the missile configuration, stabilization method, equations of motion, and engagement scenarios in Sections~\ref{missile_config}, \ref{stab_model}, \ref{EQOM}, and \ref{engagement}, respectively; these are condensed versions of the relevant sections in \cite{gaudet2020_foo}.

\subsection{Missile Configuration}\label{missile_config}

The missile is modeled as a cylinder of height  $h=1 \text{ m}$ and radius $r=0.25\text{ m}$ about the missile's body frame x-axis  with a wet and dry mass $m$ of 50 kg and 25 kg, respectively, and inertia tensor as given in Equation~\ref{eq:inertia_tensor}, where the inertia tensor principal axes correspond to the missile body frame axes. 

\begin{equation}
    \label{eq:inertia_tensor}
    {\bf J}=m\begin{bmatrix} r^2 / 2 & 0 & 0 \\ 0 & (3r^2  + h^2)/12 & 0 \\ 0 & 0 & (3r^2+h^2)/12\end{bmatrix}
\end{equation}

Four divert thrusters and 12 attitude control thrusters are positioned as shown in Table~\ref{tab:thrusters}. The attitude control thrusters operate in pairs, i.e., firing thrusters 4 and 5 cause a clockwise torque around the  missile's x-axis, whereas  firing thrusters 6 and 7 together create a counter-clockwise torque around the x-axis. Each divert thruster creates 5000 N of force, whereas the attitude control thrusters each create 125 N of force. The  thrusters can be switched on or off at the guidance frequency of 25 Hz. With a 5\% shift in the missile's center of mass (caused by fuel consumption), the torques caused by a divert thrust can be exactly cancelled by firing the appropriate attitude control thrusters. For center of mass variation less than 5\%, the attitude control thrusters will overcompensate for the torque induced by the divert thrust. The nominal (wet mass) missile center of mass is assumed to be [0,0,0] in body frame coordinates, and we define center of mass variation as a percentage of the missile dimensions, i.e., a 5\% variation would offset the center of mass by +/- 2.5 cm (0.05 * $h$/2) in the body frame $x$ direction and 1.25 cm (0.05 * $r$) in the $y$ and $z$ body frame directions. The instantaneous center of mass is as shown in Equation~\ref{eq:COM}, where $\mathbf{r}_{\mathrm{com}}(t)$ is the instantaneous center of mass at time $t$, $\mathbf{r}_{\mathrm{com}}(t_o) \in \mathbb{R}^3$ is chosen from a uniform distribution at the start of an episode within the range given in Table~\ref{tab:ic}, $f_{used}$ is the fuel used up to time $t$, and $f_{max}$ is the amount of fuel at the start of the engagement (25kg). 

\begin{equation}
	\label{eq:COM}
	\mathbf{r}_{\mathrm{com}}(t) = (\mathbf{r}_{\mathrm{com}}(t_o)) (f_{used}) / (f_{max})
\end{equation}

\begin{table}[h]
	\fontsize{10}{10}\selectfont
    \caption{Body Frame Thruster Locations.}
   \label{tab:thrusters}
        \centering 
   \begin{tabular}{c r  r  r  r  r  r  r } 
      \hline
      & \multicolumn{3}{c}{Direction Vector} & \multicolumn{3}{c}{Location} & \multicolumn{1}{c}{Rotation}  \\
       \hline
      Thruster & x  & y  & z  &  x (m) & y (m) & z (m) & Axis\\
      \hline
      1 & 0.00 & -1.00 & 0.00 & 0.00 & -0.25 & 0.00  & N/A  \\
      2 & 0.00 & 1.00 & 0.00 & 0.00 & 0.25 & 0.00 & N/A \\
      3 & 0.00 & 0.00 & 1.00 & 0.00 & 0.00 & 0.25 & N/A\\
      4 & 0.00 & 0.00 & -1.00 & 0.00 & 0.00 & -0.25 & N/A \\
      
      5 & 0.00 & 0.00 &  1.00 & 0.00 & -0.25 & 0.00 &  x  \\
      6 & 0.00 & 0.00 & -1.00 & 0.00 &  0.25 & 0.00 &  x  \\

      7 & 0.00 & -1.00 &  0.00 & 0.00 & 0.00&  0.25 &  x   \\   
      8 & 0.00 &  1.00 &  0.00 & 0.00 & 0.00& -0.25 &  x    \\  

      9 & 0.00 &  0.00 & -1.00 &  0.5 & 0.00 & -0.25 &  y \\
     10 & 0.00 &  0.00 &  1.00 & -0.5 & 0.00 &  0.25 &  y \\

     11 & 0.00 & 0.00 &  1.00 &  0.5 & 0.00 &  0.25 &  y     \\
     12 & 0.00 & 0.00 & -1.00 & -0.5 & 0.00 & -0.25 &  y    \\  

     13 & 0.00 & -1.00 &  0.00 &  0.5 & -0.25 & 0.00  &  z    \\  
     14 & 0.00 &  1.00 &  0.00 & -0.5 &  0.25 & 0.00  &  z   \\  

     15 & 0.00 &  1.00 &  0.00 & 0.5 &  0.25 & 0.00  &  z    \\ 
     16 & 0.00 & -1.00 &  0.00 & -0.5 & -0.25 & 0.00 &  z    \\ 

   \end{tabular}
\end{table}

\subsection{Stabilization Model}\label{stab_model}

Since the scale factor error compensated seeker angles $\bar{\theta}_{u}^B$ and $\bar{\theta}_{v}^B$ are in the missile body frame, which in general can be rotating, the GN\&C system must rotate them back to an inertial reference frame so that missile rotations are not confused with target maneuvers. In the following, we will refer to the inertial reference frame associated with the missile's attitude at the start of the engagement as $N'$. Specifically, we start by computing the reconstructed line of sight direction vector $\hat{\boldsymbol{\lambda}}_r^B$ as shown in Equations~\ref{eq:seeker1} through \ref{eq:seeker4}.

\begin{subequations}
\begin{align}
y &= \sin(\bar{\theta}_u^B) \label{eq:seeker1}\\
z &= \sin(\bar{\theta}_v^B) \label{eq:seeker2}\\
x &= \sqrt{1-z^2-y^2} \label{eq:seeker3}\\
\hat{\boldsymbol{\lambda}}_r^B &= [x,y,z] \label{eq:seeker4}
\end{align}
\end{subequations}

Further, we define  $\mathbf{C}_\mathrm{BN'}(\mathbf{dq})$ as the DCM mapping from the inertial frame $N'$ to the body frame given $\bf dq$. We can now define the stabilized seeker angles $\theta_u^S$ and $\theta_v^S$, and compute them as shown in Equations~\ref{eq:seeker5} through \ref{eq:seeker7}. 

\begin{subequations}
\begin{align}
\hat{\boldsymbol{\lambda}}^S &= \mathbf{C}_\mathrm{BN'}(\mathbf{dq})^T \hat{\boldsymbol{\lambda}}_r^B \label{eq:seeker5}\\
\theta_{u}^S &= \mathrm{arcsin}(\boldsymbol{\hat\lambda}^\mathrm{S} \cdot \hat{\mathbf{u}})\label{eq:seeker6}\\
\theta_{v}^S &= \mathrm{arcsin}(\boldsymbol{\hat\lambda}^\mathrm{S} \cdot \hat{\mathbf{v}})\label{eq:seeker7}
\end{align}
\end{subequations}

Since we assume the change in attitude cannot be directly measured, we must integrate the scale factor error compensated rotational velocity vector $\bar{\boldsymbol{\omega}}$ to obtain an estimate of $\bf{dq}$ parameterized as a quaternion, as shown in Equation~\ref{eq:dqi}, where $\mathbf{dq}$ is reset at the start of each episode $\mathbf{dq}_0 = \begin{bmatrix} 1 & 0 & 0 & 0 \end{bmatrix}$.  In our simulation model, we approximate this integration using fourth order Runge-Kutta integration with a 20 ms timestep.  

\begin{equation}
    \label{eq:dqi}
    \begin{bmatrix} \dot{\text{dq}{0}} \\ \dot{\text{dq}{1}} \\ \dot{\text{dq}{2}} \\ \dot{\text{dq}{3}}\end{bmatrix} = \frac{1}{2}\begin{bmatrix} \text{dq}{0} & -\text{dq}{1} & -\text{dq}{2} & -\text{dq}{3}\\ \text{dq}{1} & \text{dq}{0} & -\text{dq}{3} & \text{dq}{2}\\ \text{dq}{2} & \text{dq}{3} & \text{dq}{0} & -\text{dq}{1} \\ \text{dq}{3} & -\text{dq}{2} & \text{dq}{1} & \text{dq}{0} \end{bmatrix} \begin{bmatrix} 0 \\ \bar{\omega}_{0} \\ \bar{\omega}_{1} \\ \bar{\omega}_{2} \end{bmatrix}
\end{equation}

\subsection{Equations of Motion}\label{EQOM}

The force $\mathbf{F}_{B}$ and torque $\mathbf{L}_{B}$ in the missile's body frame for a given commanded thrust depends on the placement of the thrusters in the missile structure. We can describe the placement of each thruster through a body-frame direction vector $\mathbf{d}$ and position vector $\mathbf{r}$, both in $\mathbb{R}^3$.  The direction vector is a unit vector giving the direction of the body frame force that results when the thruster is fired.  The position vector gives the body frame location with respect to the missile centroid,   where the force resulting from the thruster firing is applied for purposes of computing torque, and in general the center of mass ($\mathbf{r}_\mathrm{com}$) varies with time as fuel is consumed. For a missile with $k$ thrusters, the body frame force and torque associated with one or more  thrusters firing is then as shown in Equations~\eqref{eq:Thruster_modela} through~\eqref{eq:Thruster_modelc}, where $T_{\mathrm{com}}^{(i)}$ is the commanded thrust for thruster $i$, $\mathbf{d}^{(i)}$ the direction vector for thruster $i$,  $\mathbf{r}^{(i)}$ the position of thruster $i$, and $\tilde{\mathbf{F}}_{B}^{(i)}$ the force contribution for thruster $i$. The total body frame force and torque are calculated by summing the individual forces and torques.

\begin{subequations}
\begin{align}
    \tilde{\mathbf{F}}_{B}^{(i)}&=\mathbf{d}^{(i)} T_{\mathrm{com}}^{(i)} \label{eq:Thruster_modela}\\
    \tilde{\mathbf{F}}_{B}&=\sum_{i=1}^{k} \tilde{\mathbf{F}}_{B}^{(i)} \label{eq:Thruster_modelb}\\
	\tilde{\mathbf{L}}_{B}&=\sum_{i=1}^{k}(\mathbf{r}^{(i)}-\mathbf{r}_\mathrm{com})\times\tilde{\mathbf{F}}_{B}^{(i)}\label{eq:Thruster_modelc}
\end{align}
\end{subequations}

The force and torque are then passed through a first order lag simulated by integrating Equations~\ref{eq:T_lag1} and \ref{eq:T_lag2} , where $\tau_{\mathrm{u}}$ is the time constant of the first order lag. This models the thruster ignition lag.

\begin{subequations}
\begin{align}
	\dot{\mathbf{F}}_{B}=(\tilde{\mathbf{F}}_{B} - \mathbf{F}_{B}) / \tau_{\mathrm{u}}\label{eq:T_lag1}\\
	\dot{\mathbf{L}}_{B}=(\tilde{\mathbf{L}}_{B} - \mathbf{L}_{B}) / \tau_{\mathrm{u}}\label{eq:T_lag2}
\end{align}
\end{subequations}

The dynamics model uses the missile's current attitude $\mathbf{q}$ to convert the body frame thrust vector to the inertial frame as shown in Equation~\eqref{eq:BtoN} where $\mathbf{C}_\mathrm{BN}(\mathbf{q})$ is the direction cosine matrix mapping the inertial frame to body frame obtained from the current attitude parameter $\mathbf{q}$.

\begin{equation}
	\label{eq:BtoN}
	\mathbf{F}_{N}=\left[\mathbf{C}_\mathrm{BN}(\mathbf{q})\right]^{T}\mathbf{F}_{B}
\end{equation}

The rotational velocities $\bm{\omega}_{B}$ are then obtained by integrating the Euler rotational equations of motion, as shown in Equation~\eqref{eq:EulerRot}, where $\mathbf{L}_{B}$ is the body frame torque as given in Equation~\eqref{eq:Thruster_modelb}, and $\mathbf{J}$ is the missile's inertia tensor. Note we have included a term that models a rotation induced by a changing inertia tensor, which in general is time varying as the missile consumes fuel.  Specifically, the inertia tensor is recalculated at each time step to account for fuel consumption, but we do not modify the inertia tensor to account for changes in the missile's center of mass.

\begin{equation}
	\label{eq:EulerRot}
	\mathbf{J}{\dot{\bm{\omega}}_{B}}=-\Tilde{\bm{\omega}}_{B}\mathbf{J}\bm{\omega}_{B}-\dot{\mathbf{J}}\bm{\omega}_B+\mathbf{L}_{B}
\end{equation}

The missile's attitude is then updated by integrating the differential kinematic equations shown in Equation~\eqref{eq:diffeqom}, where the missile's attitude is parameterized using the quaternion representation and $\bm{\omega}_{i}$ denotes the $i^{th}$ component of the rotational velocity vector $\bm{\omega}_{B}$. 

\begin{equation}
    \label{eq:diffeqom}
    \begin{bmatrix} \dot{q_{0}} \\ \dot{q_{1}} \\ \dot{q_{2}} \\ \dot{q_{3}}\end{bmatrix} = \frac{1}{2}\begin{bmatrix} q_{0} & -q_{1} & -q_{2} & -q_{3}\\ q_{1} & q_{0} & -q_{3} & q_{2}\\ q_{2} & q_{3} & q_{0} & -q_{1} \\ q_{3} & -q_{2} & q_{1} & q_{0} \end{bmatrix} \begin{bmatrix} 0 \\ \omega_{0} \\ \omega_{1} \\ \omega_{2} \end{bmatrix}
\end{equation}

The missile's translational motion is modeled as shown in Equations~\ref{eq:EQOMa} through \ref{eq:EQOMc}. 

\begin{subequations}
\begin{align}
	{\Dot{\mathbf r}} &= {{\mathbf v}}\label{eq:EQOMa}\\
	{\Dot{\bf v}} &= \frac{{{\bf F}_{N}}}{m} + \mathbf{g}_M(x_E,y_E,z_E)\label{eq:EQOMb}\\
	\Dot{m} &= -\frac{\sum_{i}^{k}\lVert{{\bf F}_{B}}^{(i)}\rVert}{I_\text{sp}g_\text{ref}} \label{eq:EQOMc}
\end{align}
\end{subequations}
Here  ${{\bf F}_{N}}^{(i)}$ is the inertial frame force as given in Equation~\eqref{eq:BtoN}, $k$ is the number of thrusters, $g_\text{ref}=9.81$ $\text{m}/\text{s}^{2}$,  $\mathbf{r}$ is the missile's position in the engagement reference frame, and $\bf{g}_M$ is the gravitational acceleration acting on the missile, with $x_E$, $y_E$, and $z_E$ the missile's coordinates in the Earth centered reference frame. 

The target is modeled as shown in Equations~\eqref{eq:TEQOMa} and \eqref{eq:TEQOMb}, where $\mathbf{a}_{\mathrm{T}_\mathrm{com}}$ is the commanded acceleration for the target maneuver, and $\mathbf{g}_{\mathrm{T}}$ the gravitational acceleration acting on the target.

\begin{subequations}
\begin{align}
	{\Dot{\mathbf r}} &= {{\mathbf v}}\label{eq:TEQOMa}\\
	{\Dot{\mathbf v}} &= \mathbf{a}_{\mathrm{T}_\mathrm{com}} + \mathbf{g}_{\mathrm{T}}(x_E,y_E,z_E)\label{eq:TEQOMb}
\end{align}
\end{subequations}

The equations of motion are updated using fourth order Runge-Kutta integration.  For ranges greater than 1000 m, a timestep of 20 ms is used, and for the final 1000 m of homing, a timestep of 0.067 ms is used in order to more accurately measure miss distance; this technique is borrowed from \cite{zarchan2012tactical:2}.

\subsection{Engagement Scenario}\label{engagement}

\begin{figure}[h]
\begin{center}
\includegraphics[width=.6\linewidth]{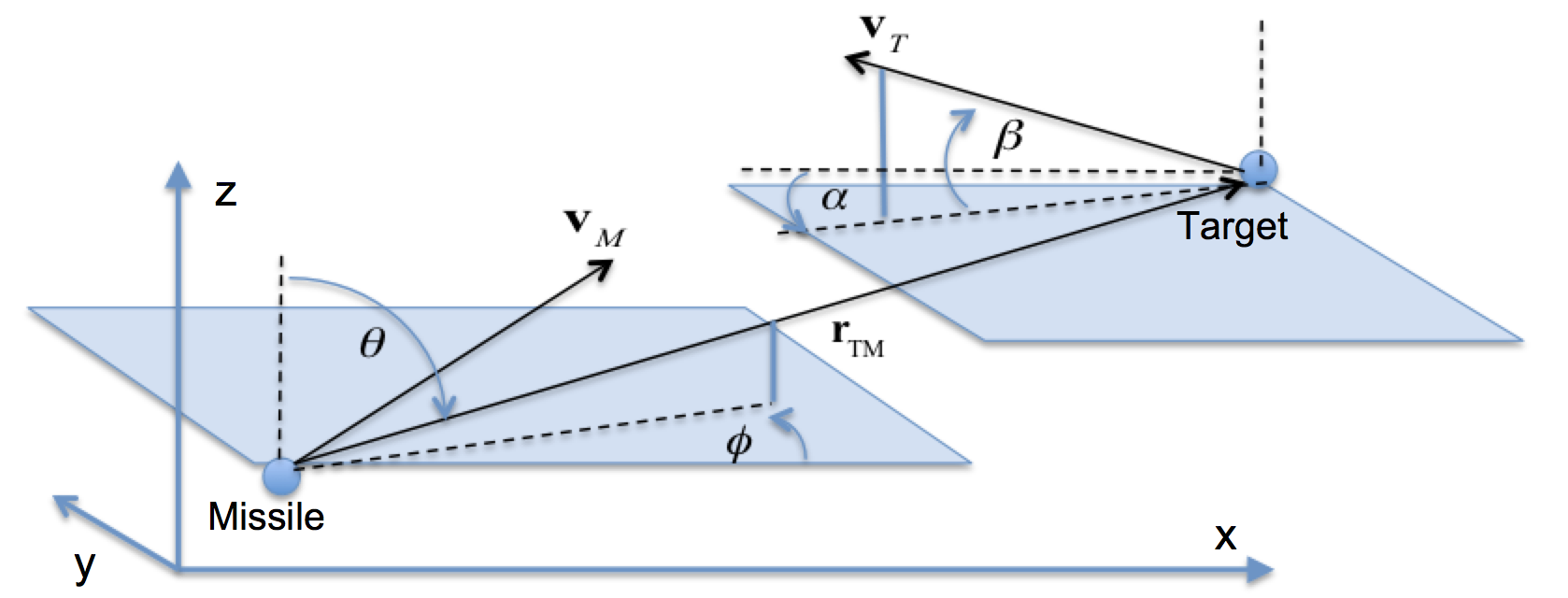}
\caption{Engagement}
\label{fig:engagement}
\end{center}
\end{figure}

The engagement is modeled as a simple skewed head-on engagement as shown in Fig.~\ref{fig:engagement}, where the collision triangle is modified to account for the gravitational field. During optimization we randomly choose between a target bang-bang and vertical-S target maneuver with equal probability, with the acceleration applied orthogonal to the target's velocity vector. The maneuvers have varying acceleration levels up to a maximum of $\mathrm{5*9.81\ m/s}^2$, and with random start time, duration, and switching time.  The range of engagement scenario parameters is shown in Table~\ref{tab:ic}.  At the start of each episode, these parameters are drawn uniformly between their minimum and maximum values.  

\begin{table}[h!]
	\fontsize{10}{10}\selectfont
    \caption{Initial Conditions}
   \label{tab:ic}
        \centering 
   \begin{tabular}{l  r  r } 
      \hline
      Parameter & min & max \\
      \hline
      Range $\|\mathbf{r}_\mathrm{TM}\|$ (km) & 50 & 55\\
      Missile Velocity Magnitude (m/s) & 3000 & 3000 \\
      Target Position angle $\theta$ (degrees) & 80 & 100 \\
      Target Position angle $\phi$ (degrees) & -10 & 10 \\
      Target Velocity Magnitude (m/s) & 4000 & 4000 \\
      Target Velocity angle $\beta$ (degrees) & -10 & 10 \\
      Target Velocity angle $\alpha$ (degrees) & -10 & 10 \\
      Heading Error (degrees) & 0 & 5 \\
      Attitude Error (degrees) & 0 & 5 \\
      Target Maximum Acceleration  $\mathrm{m/s^2}$ & 0 & 5*9.81\\
      Target Bang-Bang duration  (s) & 1 & 4 \\
      Target Bang-Bang initiation time (s) & 0 & 6 \\
      Target Barrel Roll / Vertical-S Period (s) & 1 & 5 \\
      Target Barrel Roll / Vertical-S Offset (s) & 1 & 5 \\
      \hline
      Center of Mass Variation $r_{\mathrm{com}}$  (\%) & -2.5 & 2.5\\
      Thruster Ignition Time Constant $\tau_{\mathrm{u}}$ ms & 20 & 20\\
      Angle Filter Time Constant $\tau_{\theta}$ ms & 20 & 20\\
      Seeker Angle Gaussian Noise $\sigma_{\theta}$ (rad)  & $1\times10^{-3}$ & $1\times10^{-3}$\\
      Rotational Velocity Gaussian Noise  $\sigma_{\omega}$ (rad/s) & $1\times10^{-3}$ & $1\times10^{-3}$\\
      \hline
   \end{tabular}
\end{table}

\section{Results}\label{Results}

\subsection{Baseline Performance}\label{baseline}

To establish a performance baseline, we simulate the engagement scenarios from \cite{gaudet2020_foo} without any scale factor compensation.  The statistics in the following are collected from running 5000 episodes with initial conditions randomly set as shown in Table~\ref{tab:ic}.  In Table~\ref{tab:Cases}, the "LAD" column indicates whether $\epsilon_{\theta_u}$ and $\epsilon_{\theta_u}$ are seeker angle dependent, and generated as shown in Section~\ref{LADSE} Equations~\eqref{la1} through \eqref{la3} (indicated by "Yes") or are held constant during an episode as shown in Section~\ref{LADSE} Equations~\eqref{a1} and \eqref{a2}. The PCM is trained using case~3, which is highlighted. Note that for case~6  the amplitudes of the scale factor errors are not randomized, but set to the maximum values used in case~3. 

Table~\ref{tab:baseline} gives the performance of the GN\&C system without scale factor compensation. Here we measure performance by considering the percentage of episodes that result in successful intercepts (terminal miss less than 50cm). We also include 100cm miss statistics to demonstrate performance in less demanding applications. Table~\ref{tab:baseline} also shows fuel consumption statistics. We expect fuel consumption to increase with higher scale factor errors due to excess  control activity induced by the parasitic attitude loop. Accuracy and fuel efficiency both deteriorate significantly with larger scale factor errors.  Note that in cases 0 and 1 the episodes never terminate prematurely due to a constraint violation (See Section \ref{sfcm}), but in cases~2-6 part of the performance degradation is due to constraint violations.  The constraint violations are due to abnormal control activity caused by the parasitic attitude loop described in the introduction.

\definecolor{Gray}{gray}{0.9}
\begin{table}[h!]
    \caption{Cases}
   \label{tab:Cases}
        \centering 
   \begin{tabular}{l l r r r r} 
      \hline
      Case & LAD & $A_{\theta_u}^{\text{MIN}}$ & $A_{\theta_v}^{\text{MAX}}$ & $A_{\omega}^{\text{MIN}}$ & $A_{\omega}^{\text{MAX}}$\\
      \hline
      0   & No & -1e-4 & 1e-4 & -1e-4 & 1e-4\\
      1   & No & -1e-3 & 1e-3 & -1e-3 & 1e-3\\
      2   & No & -5e-3 & 5e-3 & -5e-3 & 5e-3\\
      \rowcolor{Gray}
      3   & Yes   & 0  & 5e-3 & 0 & 5e-3\\
      4   & No & -1e-2 & 1e-2 & -1e-2 & 1e-2\\
      5   & Yes   & 0  & 1e-2 & -1e-2 & 1e-2\\
      6   & Yes   & 5e-3  & 5e-3 & 5e-3 & 5e-3\\
      \hline
   \end{tabular}
\end{table}

\begin{table}[h!]
    \caption{Performance Baseline}
   \label{tab:baseline}
        \centering 
   \begin{tabular}{l  r  r  r  r } 
      \hline
       & \multicolumn{2}{c}{Hit (\%)} & \multicolumn{2}{c}{Fuel (kg)}\\
      \hline
      Case & $<$ 100 cm & $<$ 50 cm  & $\mu$ &  $\sigma$ \\
      \hline
      0 & 100 & 98 & 10.0 & 3.1 \\
      1  & 99 & 93 & 15.6 & 2.2 \\
      2  & 90 & 72 & 16.9 & 4.5 \\
      \rowcolor{Gray}
      3  & 91 & 73 & 16.5 & 4.5 \\
      4  & 55& 37 & 21.0 & 4.4 \\
      5  & 59 & 41 & 20.6 & 4.4 \\
      6  & 60 & 45 & 20.1 & 4.9 \\
      \hline
   \end{tabular}
\end{table}
\subsection{Optimization}\label{optimization}

\begin{figure}[h!]
\begin{center}
\includegraphics[width=0.8\linewidth]{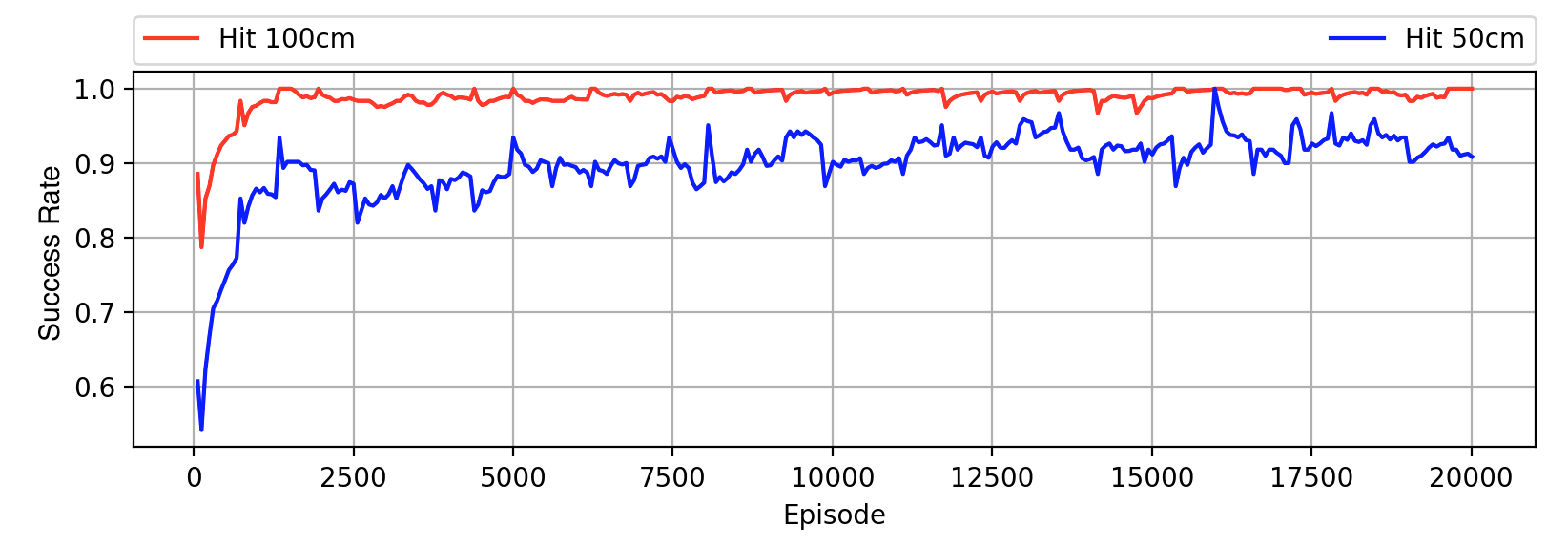}
\caption{Optimization Curve}
\label{fig:optim}
\end{center}
\end{figure}

The model is trained for 20000 episodes using the same engagement scenario that was used to establish the baseline performance, with scale factor errors set according to Table~\ref{tab:Cases}, case 3. The learning rate was set to $5\times10^{-5}$.  Fig.~\ref{fig:optim} illustrates how the fraction of successful intercepts ("Success Rate") progresses with training, with training progress measured in the number of episodes of interaction between agent and environment. The "Hit 100cm" curve illustrates training progress with a successful intercept defined as a miss distance of less than 100cm, and the "Hit 50cm" requires a miss distance of less than 50cm. Each point on the curve is calculated using statistics from the last ten rollouts (1200 episodes), and is updated every ten rollouts as  training progresses. We see that performance starts out slightly worse than the no compensation baseline, but the model learns fairly quickly, exceeding baseline performance (Table~\ref{tab:baseline} case 3) after 1200 episodes of training, and then continues to improve at a slower rate.

\subsection{Testing}\label{testing}

\begin{figure}[h!]
\begin{center}
\includegraphics[width=0.75\linewidth]{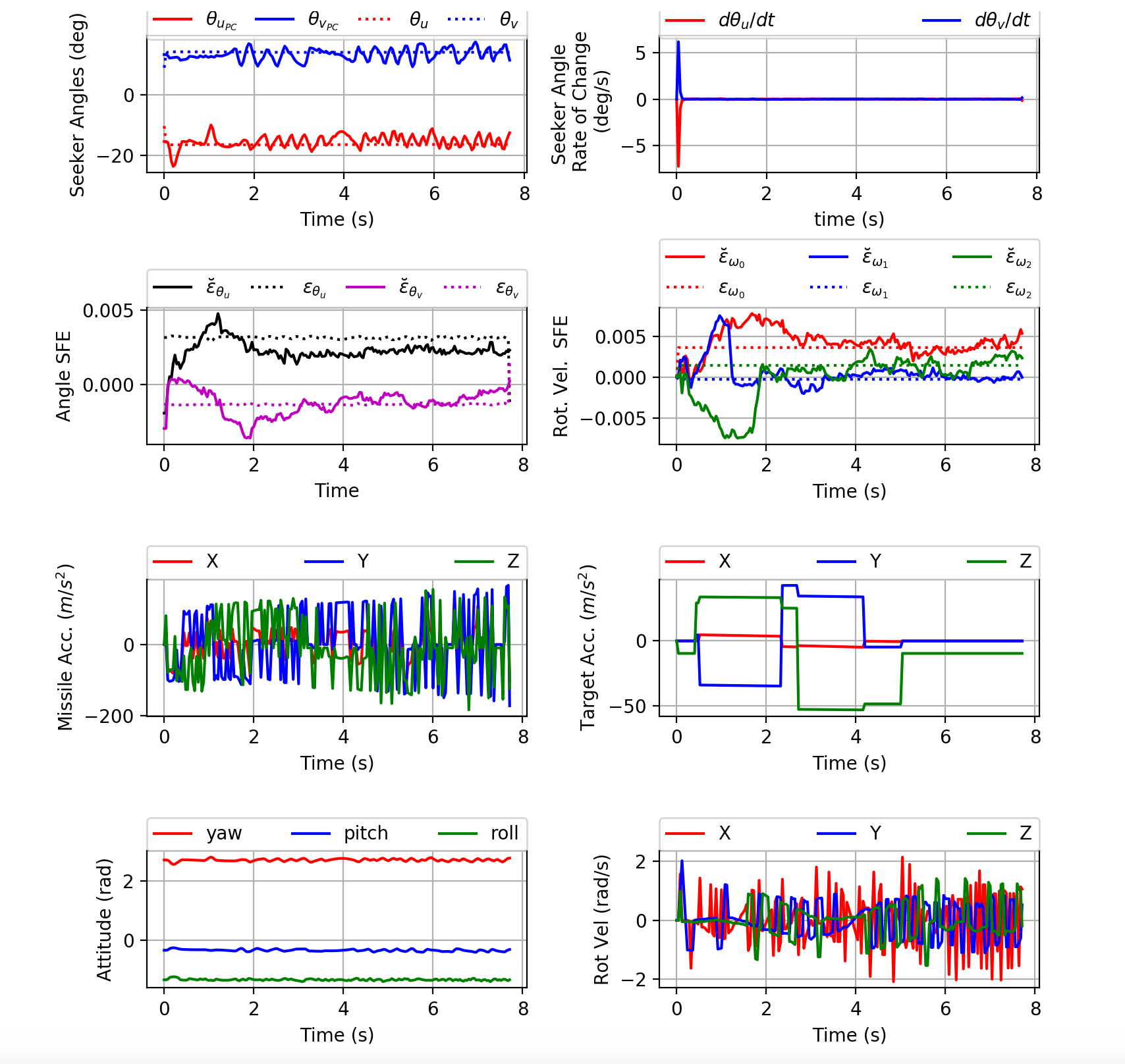}
\caption{Sample Trajectory}
\label{fig:traj}
\end{center}
\end{figure}

The optimized model was then tested for 5000 episodes under the same conditions given in Table~\ref{tab:Cases}, with results tabulated in Table~\ref{tab:comp_perf}. For cases 1 through 3 no episodes terminated prematurely due to constraint violations, but occasional constraint violations occurred for cases 4 and 5. Clearly the scale factor  compensation method is effective, giving a high probability of intercept for scale factor errors up to $1\times10^{-2}$. Figure~\ref{fig:traj} illustrates a randomly selected trajectory from case 3.  Numbering the subplots from left to right and then top to bottom, subplot 1 illustrates the unstabilized and uncompensated seeker angles $\theta_{u_\text{PC}}$ and $\theta_{v_\text{PC}}$, and the compensated and stabilized seeker angles $\theta_{u}$ and $\theta_{v}$, which are considerably smoother.  Subplot 2 shows the stabilized seeker angle rates of change, and in subplot 3 the ground truth seeker angle scale factor errors $\epsilon_{\theta_u}$ and $\epsilon_{\theta_v}$ are plotted along with their estimated values $\breve{\epsilon}_{\theta_u}$ and $\breve{\epsilon}_{\theta_v}$. Similarly, subplot 4 plots the ground truth rotational velocity scale factor error vector $\boldsymbol{\epsilon}_\omega$ and its estimated value $\breve{\boldsymbol{\epsilon}}_\omega$. We see that although the compensation is not perfect, it is sufficient to give a significant increase in kill probability as compared to the uncompensated system, particularly for the higher scale factor errors.   The remaining subplots should be self explanatory.

\begin{table}[h!]
    \caption{Performance with Scale Factor Compensation}
   \label{tab:comp_perf}
        \centering 
   \begin{tabular}{l  r  r  r  r } 
      \hline
       & \multicolumn{2}{c}{Hit (\%)} & \multicolumn{2}{c}{Fuel (kg)}\\
      \hline
      Case & $<$ 100 cm & $<$ 50 cm  & $\mu$ &  $\sigma$ \\
      \hline
      1  &  100 & 96 & 12.5 & 3.2\\
      2  &  100 & 94 & 13.0 & 3.2\\
      \rowcolor{Gray}
      3  &  100 & 94 & 13.1 & 3.3\\
      4  &  98 & 89 & 14.3 & 3.6 \\
      5  &  97 & 89 & 14.4 & 3.8\\
      6 &  97 & 88 & 14.4 & 3.8 \\
      \hline
   \end{tabular}
\end{table}

\subsection{Discussion}\label{discussion}

Using a dynamic dataset (the rollouts) for training is critical to obtaining good performance, as it  mitigates the distribution mismatch problem that can occur with static datasets \cite{ross2011reduction}. A compensation network trained using a static dataset will obtain good performance on the training set, but perform poorly when deployed.  To see why, consider that as the model improves, the compensation network's mapping  $\mathcal{G}: \mathbf{o} \mapsto \bar{\mathbf{o}}$ changes. Since $\mathbf{u} = \pi(\bar{\mathbf{o}})$, the trajectories induced by $\pi$ at the end of training will be quite different than those used to construct the dataset. Thus, the distribution of the static training dataset $[\mathbf{o}\hspace{5pt}\mathbf{u}\hspace{5pt}\boldsymbol{\epsilon}]$ may bear little resemblance to the distribution of trajectories seen during deployment. Contrast this to the case where the dataset is dynamic and implemented as a set of rollouts as discussed in Section~\ref{sfcm}.  Here, the model is updated using the most recent set of trajectories captured in the rollouts as the agent and environment interact, and when training completes, the performance of the deployed policy will be close to that of the performance on the last few training rollouts.

The network's recurrent layer, although required in a PCM, also allows the model to infer the value of signals that are only observable when considering a sequence of prediction errors and actions. In other words, the compensation method is adaptive in that it can estimate the scale factor errors in real time as the agent interacts with the environment. For example, consider the case of adapting to a constant, but unknown, rotational velocity scale factor error vector $\boldsymbol{\epsilon}_\omega$, where the network is trained in an environment that randomly sets $\boldsymbol{\epsilon}_\omega$ at the start of each episode  $\boldsymbol{\epsilon}_\omega = \mathcal{U}(\epsilon_{\omega_\text{MIN}},\epsilon_{\omega_\text{MAX}},3)$, where  $\epsilon_{\omega_\text{MIN}}$ and $\epsilon_{\omega_\text{MAX}}$ bound the scale factor errors expected during deployment. For a given rotational velocity $\boldsymbol\omega_t$ and missile inertia tensor $\mathbf{J}_t$ at time $t$, the torque $\mathbf{L}$ resulting from the missile's thrust vector will result in a new rotational velocity $\boldsymbol\omega_{t+1}$. However, the measured rotational velocity vectors $\tilde{\boldsymbol\omega}_t$ and $\tilde{\boldsymbol\omega}_{t+1}$ will be as given in Equation~\eqref{sf3}. Momentarily neglecting the Gaussian noise term, clearly the ratio of $\tilde{\boldsymbol\omega}_t$ to $\tilde{\boldsymbol\omega}_{t+1}$ depends on the ground truth value of $\boldsymbol{\epsilon}_\omega$, allowing inference of $\boldsymbol{\epsilon}_\omega$ only if a model considers at least two sequential samples of  $\tilde{\boldsymbol\omega}$.  Moreover, for the case where we do not neglect the Gaussian noise term, inference will be more robust to noise if the model considers a longer history of $\tilde{\boldsymbol\omega}$, as the impact of noise can be averaged out. It follows that an approach that does not use a recurrent network layer such as that suggested in \cite{lin2001stability} would be ineffective in this application, as the network would just learn the mean of the distribution of scale factor errors seen during training. 

Note that in the previous example, the torque $\mathbf{L}$ induced by the missile's thrust vector also depends on the evolution of the missile's center of mass during an episode.  Thus, for a given $\boldsymbol{\epsilon}_\omega$ and thrust vector, the change in the missile's ground truth rotational velocity $\dot{\boldsymbol{\omega}}$ will vary depending on the missile's current center of mass location, and $\dot{\boldsymbol{\omega}}$ will also vary with the missile's time varying inertia tensor $\bf J$. This further complicates the model's task of inferring $\boldsymbol{\epsilon}_\omega$, and requires looking at a history of observations for accurate inference. Similarly, the task of inferring  $\epsilon_{\theta_u}$ and $\epsilon_{\theta_u}$ is complicated because a component of the changes in observed seeker angles $\tilde{\theta}_u^B$ and $\tilde{\theta}_u^B$ is due to $\boldsymbol{\omega}$, but only $\tilde{\boldsymbol{\omega}}$ is observable by the PCM. 

Although we only tested the scale factor compensation method with scale factor errors up to the range $-1\times10^{-2} < \epsilon < 1\times10^{-2}$, this level of scale factor error is quite high for the demanding exoatmospheric intercept application, where small guidance system time constants are required to meet the hit to kill requirement for a maneuvering target at high closing velocities. This is discussed in more detail in \cite{gaudet2020_foo}. Moreover, in unpublished research using typical 3-DOF endoatmospheric interception engagement scenarios, we found that the method successfully compensates for radome error slopes of up to 0.15.

Other applications for this scale factor compensation method include endoatmospheric missiles with either strapdown or gimbaled seekers. For the case of gimbaled seekers, the seeker platform is mechanically stabilized using rate gyro measurements. When these rate gyro measurements are distorted by scale factor errors, the platform will no longer be perfectly stabilized, and the guidance system will be unable to completely differentiate between target maneuvers and apparent target maneuvers induced by the imperfectly stabilized seeker. Finally, note that the method is also applicable to real time rate gyro calibration in general, which is useful for a wide range of applications including space exploration.

\section{Conclusion}

We developed a method to adaptively compensate for scale factor errors in both rotational velocity and seeker angle measurements. The method uses a predictive coding model to estimate the scale factor errors, with the estimated scale factor errors then used to modify the observed seeker angles and rotational velocity vector, with the modified signals used by the missile's GN\&C system. The method is adaptive, in that it can estimate the potentially time varying scale factor errors in real time as the agent interacts with the environment. The effectiveness of the method was demonstrated in a realistic six degrees-of-freedom simulation of an exoatmospheric intercept against a maneuvering target. However, since the compensation method modifies the observations  passed to the missile's GN\&C system, it is independent of the actual GN\&C system implementation, making it applicable to a wide range of aerospace applications. Potential applications include missile's with gimbaled seekers, endoatmospheric missiles with strapdown seekers, and spacecraft. Indeed, the method could be used for general purpose real time rate gyro calibration. 

\bibliography{references}

\end{document}